# A Voice Interactive Multilingual Student Support System using IBM Watson

Kennedy Ralston
School of Computing
*Queen's University*
Kingston, ON, Canada
ralson@cs.queensu.ca

Yuhao Chen
School of Computing
*Queen's University*
Kingston, ON, Canada
14yc37@queensu.ca

Haruna Isah
School of Computing
*Queen's University*
Kingston, ON, Canada
isah@cs.queensu.ca

Farhana Zulkernine
School of Computing
*Queen's University*
Kingston, ON, Canada
farhana@cs.queensu.ca

*Abstract*—Systems powered by artificial intelligence are being developed to be more user-friendly by communicating with users in a progressively "human-like" conversational way. Chatbots, also known as dialogue systems, interactive conversational agents, or virtual agents are an example of such systems used in a wide variety of applications ranging from customer support in the business domain to companionship in the healthcare sector. It is becoming increasingly important to develop chatbots that can best respond to the personalized needs of their users, so that they can be as helpful to the user as possible in a real human way. This paper investigates and compares three popular existing chatbots API offerings and then propose and develop a voice interactive and multilingual chatbot that can effectively respond to users' mood, tone, and language using IBM Watson Assistant, Tone Analyzer, and Language Translator. The chatbot was evaluated using a use case that was targeted at responding to users' needs regarding exam stress based on university students survey data generated using Google Forms. The results of measuring the chatbot effectiveness at analyzing responses regarding exam stress indicate that the chatbot responding appropriately to the user queries regarding how they are feeling about exams 76.5%. The chatbot could also be adapted for use in other application areas such as student info-centers, government kiosks, and mental health support systems.

*Keywords—Chatbots, IBM Watson, Language Translator, Speech to Text, Text to Speech, Tone Analyzer*

I. INTRODUCTION

We have long envisioned that in the near future computers will fully understand natural language, anticipate our needs and proactively complete tasks on our behalf [1]. There is currently an increasing interest in both the academia and the industry to develop systems with natural conversation capabilities similar to humans [2]. Chatbots, also known as dialogue systems, interactive conversational agents, or virtual agents are an example of such systems used in a wide variety of applications ranging from technical support services to language learning and entertainment [3]. Chatbots are becoming a necessity in several industries [4]. A tremendous amount of investment has been made in developing conversational agents such as Apple's Siri, Microsoft's Cortana and XiaoIce, Google Assistant, Facebook Messenger bots, IBM's Watson Assistant, and Amazon's Alexa [5]. However, the development of chatbots with approximated emotional capabilities is a work in progress. Understanding complex user behaviour under various conditions and scenarios can be fundamental to the improvement of user experience for a given system [6]. To better mimic human emotional capabilities, chatbots should be able to do things beyond just understanding concepts presented in text-based questions, such as interpreting the tone of text/voice, and analyzing facial expressions [7].

Chatbots can function in an open domain such as social media platforms where the user can take the conversation anywhere or in a closed domain such as customer support or shopping assistant where the space of possible inputs and outputs is somewhat limited [8]. Chatbot models are broadly categorized as either retrieval- or generative-based. The retrieval-based models often use a repository of predefined responses and some kind of heuristics such as rule-based expression match or an ensemble of machine learning classifiers to pick an appropriate response based on the input and context [4]. Both the retrieval and generative approaches have some obvious pros and cons. Retrieval-based models don't make grammatical mistakes because their responses are from handcrafted repositories. However, they may be unable to handle unseen cases for which no appropriate predefined response exists in their repositories.

Generative-based models are smarter and can refer back to entities in the input to give the impression that the user is talking to a human. These models are likely to make grammatical mistakes especially on longer sentences and typically require huge amounts of training data. Recent breakthroughs in deep learning, largely due to public distribution of very large rich datasets, the availability of substantial computing power, and the development of new training methods for neural architectures [2], is transforming dialogue technologies. Deep learning techniques are used in both retrieval-based and generative models, but research seems to be moving into the generative direction [8]. A popular class of generative-based chatbot models are generative recurrent systems like seq2seq [9] which are rooted in language modelling and are able to produce syntactically coherent novel responses in conjunction with memory-augmented networks [10].

Despite the continuous progress in deep learning, which make chatbots quite reliable and able to provide automatic and adaptive human-like conversation behaviour, they need improvement with respect to supporting complex behaviours [4, 10]. The infinite number of topics and the fact that a certain amount of world knowledge is required to create reasonable responses makes the development of an open domain chatbot a hard problem [8]. There are several areas of cognition that must be programmed into a chatbot in order to make it more human-like, these include empathetic conversation modelling, knowledge and memory modelling, deep neural-symbolic reasoning, and modelling and calibration of emotional or intrinsic rewards reflected in human needs [5].

This study investigates the applications of chatbot related API packages offered by companies such as Google, Amazon, and IBM, and propose novel strategies for improving the interactivity of chatbots by incorporating support for various user languages and the ability to analyze user tone/mood (such as anger, sadness, etc.). These extended functionalities were achieved by incorporating IBM's Watson Tone Analyzer and Watson Language Translator APIs into the Watson Assistant API, which provides the standard chatbot base. This will expand the potential of a standard chatbot by allowing it to converse with a more diverse audience taking into account the user's emotional state in order to provide a more customized, empathetic response to fit their needs. This study is an exploration of how a chatbot that incorporates an understanding of user emotion can be significantly more helpful to users, as opposed to a chatbot that utilizes generic responses only.

The paper is organized as follows. Section II presents the research problem background, interactive chatbots use case scenario, literature review and a comparative study of relevant APIs provided by several companies in terms of their abilities to analyze tone and user traits and translate between languages effectively. Next, we present the architecture, the components description, and the implementation details of the proposed chatbot in Section III. Section IV presents the experimental results. Finally, we conclude in Section V by summarizing the contributions and listing the future work.

## II. BACKGROUND STUDY

### A. Use Case Scenario

The proposed chatbot is targeted towards students and focuses on question answering and stress management during exams. It assesses how the user is feeling about their exams, and offers advice based on analysis of the tone. The following are the functional requirements of the propose chatbot:
- Accept user input in text, facial expression, or voice formats.
- Translate the received input to the English language where necessary.
- Process the received or translated input to generate an appropriate response.
- Translate the received response back to the input or desired language where necessary.
- Deliver response to the user in the form of a printed text, gesture, or speech.

The chatbot can be extended to be used as a mental health support system for students who need support and have difficulty coping with the stress of exams. The system can also be customized for a customer service bot to respond to customer needs, queries and handle unsatisfied or angry customers methodically where humans need special training.

Next, we will provide a literature review of the state-of-the-art chatbot models and techniques that have been applied in designing more intuitive and usable dialogue systems.

### B. Literature Study

Chatbots have been developed in recent years for numerous purposes such as providing pregnant women with resources to answer their questions [4] and answering university student queries [11]. As stated in the introduction section, Chatbots models are broadly categorized as either retrieval- or generative-based. The main task retrieval-based chatbot is response selection, that aims to find correct responses from a pre-defined index. State-of-the-art examples of retrieval-based models include the multi-view response selection model that integrates information from word and utterance sequence views as proposed by Zhou et al. [12]; the sequential matching network (SMN) to tackle the issue of loss of relationships among utterances or important contextual information in multi-turn conversation as proposed by Wu et al. [13]; and a deep attention matching network proposed by Zhou et al. [15] that extends the attention mechanism of a Transformer [14] using stacked self- and cross-attention.

Rather than relying on predefined responses, the generative-models generate new responses from scratch by utilizing machine translation (from inputs to outputs) techniques. Typical examples of generative models include the data-driven response generation model in social media as proposed by Ritter et al. [16]; the context-sensitive generative model as proposed by Sordoni et al. [17]; the Maximum Mutual Information (MMI) as the objective function in neural generative models as proposed by Li et al. [18], and recently, the TransferTransfo model which combine transfer learning based training scheme and the Transformer [14] model as proposed by Wolf et al. [19].

Models of user reactions, responses and emotions can aid in the design of more intuitive and usable dialogue systems [6]. Hasegawa et al. [20] utilized Twitter posts to investigate the task of estimating the emotion of a speaker or writer from their utterance or writing and then proposed techniques for generating an appropriate response that elicits a specific emotion in the addressee's mind. Recently, Lubis et al. [21] proposed and evaluated a novel neural network architecture for an emotion-sensitive neural chat-based dialogue system, optimized on the constructed corpus to elicit positive emotion.

Next, we present a comparison of IBM, Amazon, and Google's API offerings in relation to natural language processing and creation of a chatbot.

TABLE I. COMPARISON OF SOME MAJOR CHATBOT API OFFERINGS

| Features | IBM | Amazon | Google |
|---|---|---|---|
| API | *Watson Assistant* | *Lex* | *Dialogflow* |
| Language Translation | Uses Watson Translate which supports language auto-detection and translation between 24 languages | Uses Amazon Translate which supports language auto-detection and translation in 25 languages | Uses Google Translate which supports language auto-detection and translation in more than 100 languages |
| Personality Analysis | Uses Watson Personality Insights which offers the ability to infer personality traits | Uses Amazon Comprehend service which offers insights derivation such as sentiment | Uses Google Natural Language API, which offers insights derivation such as sentiment |
| Tone Analysis | Uses Watson Tone Analyzer which offers the capability to detect the tone and emotion indicated by written text | Uses Amazon Comprehend service (see above). However, unable to provide specific the tone in terms of mood | Uses Google Natural Language API (see above). Also, unable to provide specific the tone in terms of mood |

### C. Comparative Study

IBM, Amazon, and Google are among the major platforms that provide APIs for the development of chatbots. IBM provides a chatbot API called Watson Assistant[1] which comes pre-trained with industry-relevant content to respond to user queries. The system can ask for clarity or for the user to rephrase when it doesn't understand a query. It is free up to the first 10,000 messages per month. Amazon provides a chatbot API called Lex[2] to build conversational interfaces for applications. Lex supports speech recognition and natural language processing. It offers 10,000 text requests and 5,000 speech requests per month for free for the first year. Google provides a chatbot API, called Dialogflow[3], which allows users to build engaging voice and text-based conversational interfaces. Dialogflow incorporates Google's machine learning expertise and products such as Google Cloud Speech-to-Text, and offers a free standard version to try out for small to medium businesses. Table 1 provides a summary of these three APIs based on their language translation, personality and tone analysis functionality. Though IBM Watson lacks slightly in terms of language translation when compared to other services, it offers the best options for analyzing textual tone in terms of mood. Therefore, we chose the IBM Watson suite to create an empathetic chatbot for analyzing user mood.

---
[1] https://www.ibm.com/cloud/watson-assistant
[2] https://aws.amazon.com/lex
[3] https://dialogflow.com

In order to predict user behaviour, it is helpful to create a profile of user's personality. The words one uses in everyday life can be used to estimate personality traits and current mood. Predicting user behaviour based on this knowledge is useful to create a chatbot because it allows the chatbot to more efficiently respond to the needs of the users [6].

IBM Watson offers two services, Tone Analyzer and Personality Insights, which allow for such prediction based on the analysis of users' text-based responses with a chatbot. Watson Personality Insights returns users' personality traits based on an analysis of their social media messages and other online data. Watson Tone Analyzer can detect users' current emotions (mainly anger, sadness, fear, joy, and disgust) and additionally can conclude their writing styles based on their text interactions (e.g. confident). It achieves this using three major models (i) personality trait model (in which the big five traits are agreeableness, conscientiousness, extraversion, emotional range, and openness), (ii) needs model (which describes the types of aspects of a product that can appeal the most to a user), and (iii) values model (factors that have an important impact on users' decision making).

*D. Related Work*

Li et al. [22] utilized IBM Watson Language Translator API to implement an automated invoice processing system for a global business that provides secure language translation between multiple languages. The system also offers a feedback system to periodically rate how accurately users' words or sentences are translated and are useful in improving the language model.

Marbouti et al. [23] used IBM Tone Analyzer to analyze public tweets to find relevant information that could be helpful during an emergency situation. The Tone Analyzer was incorporated into this program in order to judge the emotional/social context of a given tweet. Using this data along with the information about the user and the receiver(s) of the tweet, an analysis can be done to filter and sort tweets to extract useful information in the event of an emergency situation. This is a relevant use case which suggests the application of IBM Tone Analyzer to our proposed chatbot.

Pabón et al. [24] used IBM Watson Personality Insights service to examine online messages between developers to get an idea of how their personality is reflected in these messages, and the relationship between their predicted social aspects and the technical aspects of their work. This study suggests the use of Personality Insights for the improvement of customer service. Herzig et al. [25] used Watson Personality Insights to predict the satisfaction of a customer after an interaction with a customer service agent via social media. This study could be extended to target students' stress management during exams.

Packowski and Lakhana [26] utilized IBM's services to build a chatbot that could effectively respond to customer service requests. IBM's services that were used include Natural Language Classifier (NLC) for classifying chat messages into predetermined groups, Natural Language Understanding (NLU) for extracting keywords and entities, Language Translator for identifying what language chat messages have been written in, Watson Knowledge Studio for creating a custom language model, and Tone Analyzer for identifying the mood of a chat message. The authors shared the following experiences. (i) Use historical data to guide solution design choices rather than making assumptions about how people use to chat. (ii) Beyond quantity, factors such as the quality of the training data and the nature of the classes influence how much training data is needed to train a well-performing NLC. Use of an evaluation set can help identify when an NLC's performance improvement levels off, in which case, more effort spent on data collection and training may not be worthwhile. (iii) An NLC can reliably be created using the same training set. (iv) Do not assume that an NLC trained for one context will successfully classify data from another context. (v) It is a good idea to start simple, and then add more complex components as needed. This study is similar to ours. In addition, we are using IBM's Chatbot API and Watson Personality Insights.

One specific chatbot application that served as a source of inspiration for our study was the chatbot system created by Fleming et al. [11] that streamlined student course requests at the University of Queensland in Australia. The chatbot was developed using Amazon Lex and its goal was to support students from many different backgrounds by recognizing user query intent and providing a response using a database of prior student queries. The authors reported that the chatbot was successful in answering simple student queries, specifically focused on allowing students to request assignment extensions by providing their course code, the name of the assignment for which the extension was requested, and the reason for the extension.

In this study, we are leveraging IBM Watson Assistant, Tone Analyzer, Text to Speech, Text to Speech, and Language Translator to develop our proposed chatbot. Ultimately, the goal of this study is to integrate the chatbot capability with a robot that has speech recognition capabilities. Speech recognition allows people to communicate more effectively with applications and allows greater accessibility for a wider range of users who may not be able to use the general mouse/keyboard method of computer interaction. Additionally, it allows users to interact in a more natural way which allows for ease in multi-tasking [27]. In order to fulfill the needs of users, a system should be intuitive and natural to use, and speech recognition tools play an important role in this accessibility. Next, we present the architecture and components description of the proposed chatbot.

III. PROPOSED CHATBOT

The general architecture of the proposed voice interactive multi-lingual chatbot based on the use case scenario described in Section II-A is shown in Fig.1.

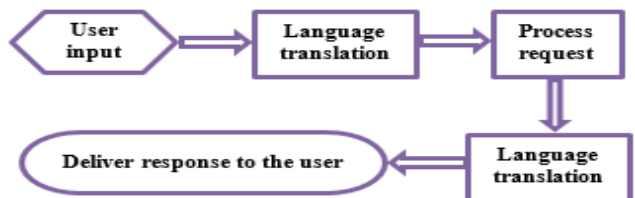

Fig. 1. General architecture of the proposed chatbot.

As stated in Section II-D, we are leveraging IBM Cloud tools including Watson Assistant, Tone Analyzer, Speech-to-

Text, Text-to-Speech, and Language Translator for developing our proposed chatbot. Watson Assistant makes it easier to build, test, and deploy a bot or virtual agent across mobile devices, messaging platforms like Slack, or even on a physical robot. Fig. 2 shows the architecture of the proposed chatbot with specific components.

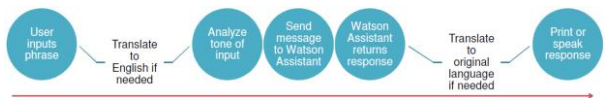

Fig. 2. General architecture of the proposed chatbot.

The proposed chatbot is an integral part of our intelligent interactive robotic system as shown in Fig. 3. The intelligent robotic system is clearly separated into two main parts namely dynamic face recognition and voice communication. In this paper we illustrate the voice communication part.

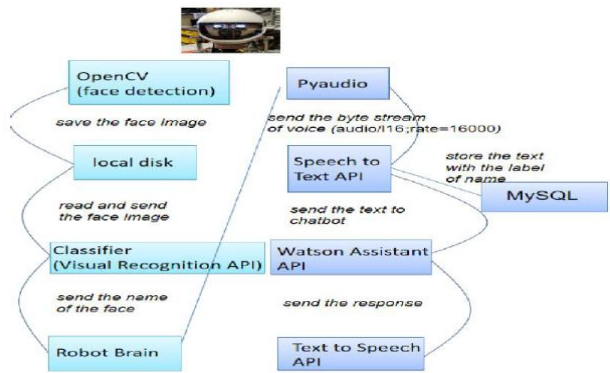

Fig. 3. Intelligent interactive robotic system.

The chatbot is targeted towards student users and focuses on stress management during exams. It assesses how the user is feeling about their exams, and then offers advice based upon its analysis of their tone. Next, we present the description of the implementation of the proposed chatbot.

We implemented the proposed chatbot on a Windows 10 Home operating system with 8.00 GB RAM, an Intel Core i3-2370M 2.40 GHz CPU, and approximately 7.7 GB storage. We used the Watson Developer Cloud SDK for Python. Speech input/output was handled with the use of the PyAudio library for Python. The following are the specific APIs that were utilized:
- IBM Watson Assistant
- IBM Watson Tone Analyzer
- IBM Watson Language Translator
- IBM Watson Text-to-Speech
- IBM Watson Speech-to-Text

Using these tools, we implemented a chatbot that takes input from student users and provides advice on how to deal with exam stress based upon how the user is feeling. The chatbot can respond to the user in about 25 languages.

### A. Data

In order to implement a chatbot that can truly help student users, a knowledge base needs to be used or developed and be filled with accurate information. We utilize results from various mental health research studies as shown in TABLE II, to develop a knowledge base for coping with exam stress and anger.

TABLE II. MENTAL HEALTH STRATEGIES TO COPE WITH EXAM STRESS

| Ways of coping with exam stress[4,5] | Ways of coping with anger around exam season[6] |
|---|---|
| Create a study schedule | Take some time to reflect |
| Get enough sleep | Take a walk/run to get some exercise |
| Eat healthy food | Express your anger |
| Drink water | Don't hold grudges |
| Take breaks for movements | Practice relaxation skills |
| Avoid distractions during study | Talk to someone you trust |
| Reward yourself when achieving study goals | Meditate |
| Talk to someone you trust | Identify possible solutions |

A selection of these solutions was implemented in the proposed chatbot to provide the best advise depending on users' emotional states to cope with the stress of exam season.

### B. Implementation Details

The first step in creating a chatbot using Watson Assistant is to define a skill which includes the data and logic that will be used by the assistant to communicate with users. A skill has three components intents, entities, and dialogs.
- An intent is a representation of the purpose of what the user is inputting into the chatbot (e.g. greetings or help requests). In Watson, it is identified by a prefix symbol '#'. The skill can be trained to recognize an intent by providing it with many user input examples and showing which manufactured intent is the appropriate mapping for each input. The intents that were used in this study was adopted from the IBM Watson Assistant Content Catalog, which provides commonly used sample intents.
- Entities are specific nouns that are important to the user's intent, such as a specific name or location. In the Watson system, it is prefixed with the "@" symbol. In this study, we only used one entity "@yesno", which simply represents whether a user answers a question with "yes" or "no". Entities in Watson Assistant also allow for the inclusion of synonyms, so that if a user says something such as "yeah" instead of "yes" in response to a question, the entity will still be interpreted correctly.
- A dialog is the organization of the flow of conversation to provide the appropriate responses to intents and entities. Each dialog node has some logical criteria that must be fulfilled in order to trigger the appropriate responses. Dialog nodes are also permitted to have children, to advance the depth of dialog on one topic. For example, Fig. 4 is a "Welcome" dialog node with children, the hierarchy of the child nodes is displayed as a reference for the options for possible user responses to the request "How are you feeling about exams?" as the user is being greeted. The user response to this question will be analyzed based on tone, and then will trigger one of these three responses. If the user's emotion is detected as fear or sadness, then the "Stressed about Exams" node will be triggered. If the user's emotion is detected as joy, then the "Feeling good about Exams" node will be triggered. If the user's emotion is detected as anger, then the "Angry about Exams" node will be triggered. Each node offers a response that is appropriate for the user's mood regarding exams.

---

[4] http://www.wrha.mb.ca/healthinfo/healthheadlines/2017/170224-managing-stress-and-anxiety.php
[5] https://www.timeshighereducation.com/student/advice/how-deal-exam-stress#surveyanswer
[6] https://www.mayoclinic.org/healthy-lifestyle/adult-health/in-depth/anger-management/art-20045434

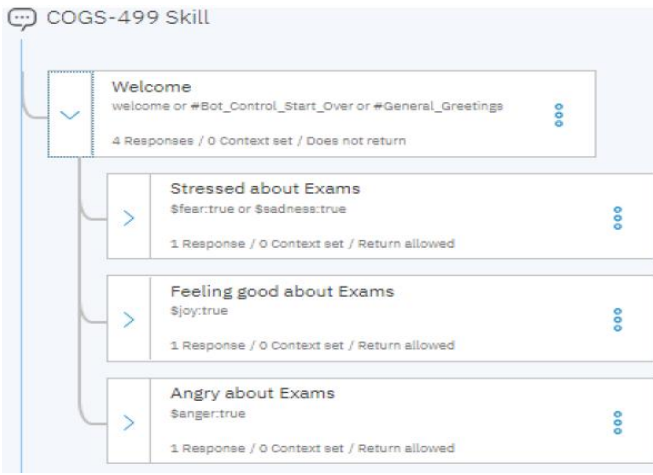

Fig. 4. "Welcome" dialog node with users.

In Fig. 5, the details of the "Stressed about Exams" node are displayed. The response given in this scenario is sympathetic and offers solutions for the user to use help cope with anxiety (which encapsulates the fear and sadness emotions). The user can then reply with some variation of "yes" if they want to hear more about dealing with stress, and some variation of "no" if they do not. Their response will trigger the "@yesno" entity discussed earlier, which will result in the correct resulting node being triggered to either tell the user more about stress relief techniques or wish them well on their exams and bring the conversation regarding stress relief to an end.

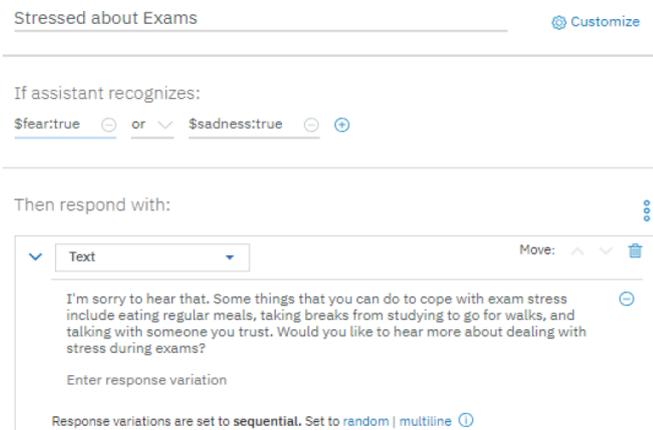

Fig. 5. "Stressed about Exams" node.

After these components are created, the skill is associated with an "Assistant" chatbot. Only one skill was necessary for the proposed chatbot, as its skill was able to contain all of the intents for our implementation.

The Watson Developer Cloud software developer kit for Python was used to connect the online Watson Assistant developer interface with the Watson Language Translation and Tone Analyzer APIs. The first step in the process is to accept the user's input for which we provide two alternatives. The user can either input queries via text or via voice using the PyAudio Library and the Watson Speech to Text API. The user's input is then processed by the Watson Language Translator API to determine the language of the input. If the language is not English, then the input is translated into English. This is necessary because the Watson Assistant intents are coded in English, and so we can only match English inputs with the appropriate intent. Once the input is translated, it is passed through the Watson Tone Analyzer API to assess the tones in the text. The processed input is then sent to the Watson Assistant API with the tone attached as a context variable. This context variable and text content are used to trigger the correct response node using Watson Assistant's logic. Once the appropriate response is chosen, it is translated back into the user's original input language, if it is not English, and is then both printed on the screen and spoken out loud to the user with the help of the Watson Text-to-Speech API.

IV. EXPERIMENTS AND RESULTS

A. Results and Evaluations

We evaluated the chatbot by examining its response to a variety of user queries. For example, we created a set of likely responses to the user query "How are you feeling about exams?" The responses (see TABLE III) were generated based on a survey of 20 current university students using Google Forms. The results of measuring the effectiveness of the chatbot at analyzing responses regarding exam stress are shown in TABLE IV. 76.5% of the time the chatbot responded appropriately to the user's query regarding how they are feeling about exams. A few times when the chatbot did not respond correctly were a) when appropriate emotion could not be detected in the user's query, or b) the user's input phrase resulted in too many analyzed emotions, and the chatbot was not able to categorize the question to generate the appropriate response, though this only happened in one case (Trial 3). One important challenge associated with this model of a chatbot is that the user can respond to the question "how are you feeling about exams?" with something completely unrelated, and the chatbot will still respond based on the emotion of what the user is saying.

TABLE III.  EXAM STRESS SURVEY RESPONSES

| Trial # | Response | # of responders |
|---|---|---|
| 1 | "Stressed" | 3 |
| 2 | "I am stressed" | 2 |
| 3 | "Bad" | 1 |
| 4 | "Stressed as per usual" | 1 |
| 5 | "I like exams better than classes" | 1 |
| 6 | "I'm feeling pretty good" | 1 |
| 7 | "I feel fine about exams as long as I study" | 1 |
| 8 | "I hate exams" | 1 |
| 9 | "Somewhat stressed" | 1 |
| 10 | "Very stressed" | 1 |
| 11 | "I think I will be ok" | 1 |
| 12 | "I hope I do well" | 1 |
| 13 | "I dislike exam season" | 1 |
| 14 | "I want it to be over" | 1 |
| 15 | "I just want them to be over" | 1 |
| 16 | "I'm hoping everything will go well" | 1 |
| 17 | "A bit stressed" | 1 |

TABLE IV.  CHATBOT EFFECTIVENESS AT ANALYZING RESPONSES REGARDING EXAM STRESS

| Trial # | Emotion | Prediction | Correct? | Why |
|---|---|---|---|---|
| 1 | Fear/Sadness | Fear | Yes | |
| 2 | Fear/Sadness | Fear | Yes | |
| 3 | Anger | Anger, Fear, Sadness | No | No response |
| 4 | Fear/Sadness | Fear, tentative | Yes | |
| 5 | Joy | Joy, tentative | Yes | |
| 6 | Joy | Joy, tentative | Yes | |
| 7 | Joy | Joy, analytical | No | Didn't understand |
| 8 | Anger | Anger | Yes | |

| 9 | Fear/Sadness | Fear, tentative | Yes | |
| 10 | Fear/Sadness | Fear, confident | Yes | |
| 11 | Joy | Analytical | No | Didn't understand |
| 12 | Joy | Joy | Yes | |
| 13 | Anger | Anger | Yes | |
| 14 | Fear/Sadness | Sadness | Yes | |
| 15 | Fear/Sadness | Sadness, tentative | Yes | |
| 16 | Joy | Joy, tentative | Yes | |
| 17 | Fear/Sadness | None | No | Didn't understand |

## V. CONCLUSIONS

Chatbots have become extremely popular in recent years due to the advancements in machine learning and other digital technologies. This paper investigates the applications of chatbot related API packages offered by Google, Amazon, and IBM and proposes a chatbot system for student support. Our comparative study indicates that despite the limitations in terms of language translation when compared to other services, IBM Watson offered the best options for specifically analyzing textual tone insights in terms of mood. Our proposed chatbot supports multiple user languages and has the ability to analyze user tone/mood (such as anger, sadness, etc.). It uses IBM's Watson Tone Analyzer and Watson Language Translator APIs into the Watson Assistant API, which provides the standard chatbot base. We utilize results from various mental health research studies to develop a knowledge base for coping with exam stress and anger and then evaluated the chatbot by examining its response to a variety of input queries. The results indicate that 76.5% of the time the chatbot responds appropriately to the user queries.

One of the limitations of Watson Assistant APIs is the need to manually create the question-answer bank for all possible questions and the associate them with intents in the chatbot's skill definition to respond appropriately to user requests. In our ongoing work we are looking into developing deep learning models to automate this process of finding the answers. Watson Language Translator API sometimes incorrectly identified languages due to irregular capitalization. Other advanced language translation APIs such as Google Translate can be explored to solve this problem. Finally, we recommend incorporating other APIs such as Google Maps to answer user inquiries about locations with a view to creating an intelligent and cognitive interactive robotic system.